# A Twitter Tale of Three Hurricanes: Harvey, Irma, and Maria

**Firoj Alam**
Qatar Computing Research Institute
Hamad Bin Khalifa University
Doha, Qatar
falam@hbku.edu.qa

**Ferda Ofli**
Qatar Computing Research Institute
Hamad Bin Khalifa University
Doha, Qatar
fofli@hbku.edu.qa

**Muhammad Imran**
Qatar Computing Research Institute
Hamad Bin Khalifa University
Doha, Qatar
mimran@hbku.edu.qa

**Michael Aupetit**
Qatar Computing Research Institute
Hamad Bin Khalifa University
Doha, Qatar
maupetit@hbku.edu.qa

**ABSTRACT**

People increasingly use microblogging platforms such as Twitter during natural disasters and emergencies. Research studies have revealed the usefulness of the data available on Twitter for several disaster response tasks. However, making sense of social media data is a challenging task due to several reasons such as limitations of available tools to analyze high-volume and high-velocity data streams. This work presents an extensive multidimensional analysis of textual and multimedia content from millions of tweets shared on Twitter during the three disaster events. Specifically, we employ various Artificial Intelligence techniques from Natural Language Processing and Computer Vision fields, which exploit different machine learning algorithms to process the data generated during the disaster events. Our study reveals the distributions of various types of useful information that can inform crisis managers and responders as well as facilitate the development of future automated systems for disaster management.

**INTRODUCTION**

Three devastating natural disasters in 2017, namely Hurricane Harvey, Hurricane Irma, and Hurricane Maria, caused catastrophic damage worth billions of dollars and numerous fatalities, and left thousands of affected people. During such life-threating emergencies, affected and vulnerable people, humanitarian organizations, and other concerned authorities search for information useful to prevent a crisis or to help others. For instance, during disasters, humanitarian organizations and other government agencies, public health authorities, and military are tasked with responsibilities to save lives and reach out to people who are in need of help (Gralla et al. 2013). These formal response organizations rely on timely and credible information to make timely decisions and to launch relief efforts. The information needs of these stakeholders vary depending on their role, responsibilities, and the situation they are dealing with (Vieweg, Castillo, et al. 2014). However, during time-critical situations, the importance of timely and factual information increases, especially when no other traditional information sources such as TV or Radio are available (Vieweg 2012; Castillo 2016).

The growing use of Information and Communication Technologies (ICT), mobile technologies, and social media platforms such as Twitter and Facebook has provided easy-to-use and effective opportunities to the general public to disseminate and ingest information. Millions of people increasingly use social media during natural and man-made disasters (Castillo 2016; Hughes and Palen 2009; Purohit et al. 2014). Research studies have demonstrated the usefulness of social media information for a variety of humanitarian tasks such as "situational awareness" (Vieweg 2012; Starbird et al. 2010). Although, information available on social media could be useful for response agencies, however, making sense of it under time-critical situations is a challenging task (Hiltz and Plotnick 2013). For instance, due to high-volume and high-velocity of social media data streams, manual analysis of thousands of social media messages is impossible (Hiltz, Kushma, et al. 2014; Ludwig et al. 2015).

Making sense of social media data to aid responders involves solving many challenges including parsing unstructured and brief content, filtering out irrelevant and noisy content, handling information overload, among others. Over the last few years, a number of Artificial Intelligence (AI) techniques and computational methods have been proposed to process social media data for disaster response and management. These techniques aim to solve various challenges ranging from information filtering, overload, and categorization to summarization of information (Castillo 2016; Imran, Castillo, Diaz, et al. 2015).

This work presents an extensive analysis of the Twitter data collected during the three disasters, namely Hurricanes Harvey, Irma, and Maria. We employ a number of state-of-the-art AI techniques ranging from unsupervised to supervised learning for the analysis of these real-world crisis events. The techniques used in this paper have a range of applications in the field of crisis response and management. Specifically, we perform sentiment analysis to determine how people's thoughts and feelings change over time as disaster events progress. To help concerned authorities to quickly sift through big crisis data, we use topic modeling techniques to understand different topics discussed during each day. To help humanitarian organizations fulfill their specific information needs, we use supervised classification techniques to classify both textual and imagery content into humanitarian categories. We also make our data available at the CrisisNLP[1] repository for researchers and practitioners to advance research in this field. We publish tweet ids and a tool to download full tweet content from Twitter.

The rest of the paper is organized as follows. In the next section, we present objectives of our analysis followed by the literature review section. Next, we describe our data collection details. Then, we present our experiments, analyses, and a discussion of the results. Finally, we conclude the paper in the last section.

**OBJECTIVES OF THE ANALYSIS**

Depending on their role and capacity, the information needs of formal response organizations and other humanitarian non-governmental organizations (NGOs) vary. Moreover, local government departments such as police, firefighters, and municipality among others, seek information that is aligned with their response priorities as well as the given crisis situation, its context, severity, and evolution over time. For instance, many humanitarian organizations seek high-level information about a crisis situation such as the scale of the disaster event, affected population size in the disaster zone, overall economic impact, urgent needs of the affected people such as food, water, and shelter. In contrast, other organizations such as police forces seek information concerning individual emergency situations such as reports of trapped people that need to rescue, or injured people that need urgent medical assistance, etc. Such cases require immediate attention of the concerned authorities. These varying information needs of different humanitarian stakeholders can be classified into two main categories: (i) information needs that help authorities understand the "big-picture" of a situation, i.e., "situational awareness", and (ii) information needs that help authorities to launch a rapid response to an emergency situation i.e.,"actionable information".

In this work, we aim to analyze the data collected during three natural disasters to understand different types of information available on social media based on "situational awareness" and "actionable information". We perform a multidimensional analysis of both textual as well as imagery content on social media.

**Textual content analysis**

We seek to gain an understanding of the textual information posted on social media during disasters from different angles. With a target to fulfill both situational awareness and actionable information needs of different humanitarian organizations, in this work we employ several state-of-the-art AI techniques to analyze and understand useful information for humanitarian decision-makers and responders while filtering out irrelevant and unimportant information to reduce information overload burden on responders.

*Targeting specific information needs using supervised classification*

Many humanitarian organizations have predetermined information needs. For example, United Nations (UN) humanitarian organizations use a cluster coordination approach in which different organizations focus on different humanitarian tasks[2] (Vieweg, Castillo, et al. 2014). For instance, World Health Organization (WHO) focuses

---

[1] http://crisisnlp.qcri.org/
[2] https://www.unocha.org/legacy/what-we-do/coordination-tools/cluster-coordination

on health, United Nations Children's Fund (UNICEF) on children and education, and World Food Programme (WFP) on food security. In this work, we take a longitudinal approach to cover different information needs of UN organizations. First, we define a taxonomy of information needs learned from previous studies (Imran, Castillo, Diaz, et al. 2015), and then use supervised machine learning techniques to automatically categorize the collected data into the predefined categories. The categories included in our taxonomy are as follows:

*Taxonomy representing several humanitarian information needs:*

- **Injured or dead people:** corresponds to the reports of injured people and fatalities due to the disaster.
- **Infrastructure and utility damage:** corresponds to the reports of damage to infrastructures such as buildings, bridges, roads, houses and other utility services such as power lines and water pipes.
- **Caution and advice:** messages that contain warnings, cautions, and advice about the disaster that could be useful for other vulnerable people or humanitarian organizations.
- **Donation and volunteering:** corresponds to the messages containing requests for donations of goods, money, food, water, shelter, etc. and/or messages containing donation offers.
- **Affected individual:** corresponds to the reports of affected people due to the disaster.
- **Missing and found people:** corresponds to the reports of missing or found people due to the disaster.
- **Sympathy and support:** corresponds to the messages that show any type of sympathy or support towards the victims of the disaster.
- **Personal:** corresponds to the personal updates that are mostly useful for the author's family and friends, but probably not for humanitarian organizations.
- **Other useful information:** corresponds to the messages which do not belong to any of the above categories but are still useful and potentially important for humanitarian organizations.
- **Irrelevant or not related:** corresponds to the messages which are not in English or not relevant to the disaster or irrelevant for humanitarian response.

Topics of discussion on social media during different disasters vary and even within an event topics change rapidly (Imran, Mitra, and Srivastava 2016). One factor that influences a change in the discussion is the disaster phase. For instance, if we divide a disaster event into three phases: *"Pre"* (i.e., the time period between the detection and impact of a disaster), *"During"* (i.e., the time period during the impact), and *"Post"* (i.e., the time period after the impact), then the discussions during these three phases are expected to be different. Another factor that might cause a change in the topic of discussion is varying aid needs of affected people. To understand the temporal variance between the different informational categories in the taxonomy, we aim to investigate the distribution of the classified messages over time.

*Learning concerns and panics of people*

Determining the sentiment of people during disasters and emergencies can help understand people's concerns, panics, and their emotions regarding various issues related to the event. It also helps responders establish stronger situational awareness of the disaster zone (Caragea et al. 2014; Nagy and Stamberger 2012). To establish such an understanding, we aim to perform the sentiment analysis on the collected data. With this analysis, we expect to find issues that cause anger and negative sentiment among affected people and outsiders. Humanitarian organizations can see this as a tool to keep an eye on public sentiment to find critical issues affecting large populations and plan their response in a timely manner.

*Tracking incidents using topic modeling*

Large-scale disasters that are either too severe (e.g., intense earthquakes) or long-running (e.g., wars, conflicts) usually cause many small-scale incidents, which are troublesome to small communities or a limited number of people. Examples of such small-scale events include "airport shutdown due to an earthquake", "school closures due to hurricane warnings", etc. Many of such events are hard to anticipate by humanitarian organizations. Therefore, responders are usually not well-prepared to handle them. Timely identification of small-scale events after a big disaster can help humanitarian responders launch timely response to help those who are in need or address the issue. For this purpose, we introduced the "Other useful information" category in the taxonomy described above. This category contains messages that do not belong to any other informative categories in the taxonomy, but important information potentially useful for humanitarian responders. We expect small-scale issues and incidents to appear in this category. Since the types of information present in the messages that belong to the "Other useful information" category are not known, we cannot use supervised machine learning techniques to understand what are those incidents or topics of discussion during a disaster. Instead, similar to (Imran and Castillo 2015), we use a state-of-the-art unsupervised machine learning technique called Latent Dirichlet Allocation (LDA) (Blei et al. 2003) to identify the latent events or topics in the "Other useful information" category.

*Identifying dominant entities using named-entity recognition*

Rapidly assessing a situation is critical for effective disaster response. Named entities such as the names of persons, organizations, and locations used in text messages provide ways to understand them better. Three typical entities have been recognized as fundamental elements in Natural Language Processing (NLP) (Finkel et al. 2005): "Persons", "Organizations" and "Locations". Among other ways, finding entities could help crisis managers rapidly sift through thousands of messages while discarding noise. For instance, a location unrelated to the event or the name of a past event can be used to filter out all messages that mention them. The name of a well-known organization (e.g., a non-governmental organization (NGO), a government agency or an established media corporation) mentioned in a message makes this message more trustworthy than if delivered by an unknown source. The location name of a specific street, bridge, park or river can help managers to send a rescue team to the right place. The name of a person can be used to identify a local contact or understand that an important person is missing. We use the Stanford Named-Entity Recognizer (Finkel et al. 2005) to extract the top 10 most frequent persons, organizations, and locations from our data.

**Multimedia content analysis**

Capturing the moment via images or videos, and sharing them online has already become a usual habit for many social media users. Thanks to this new phenomenon, social media users can easily share much more information in a much more effective way than just typing up text messages all the time to share their feelings and opinions. Therefore, analysis of this multimedia content (i.e., images and videos) bears significant potential, especially in the context of crisis response and management. For instance, an image can provide more information about the severity and extent of damage caused by a disaster, more detailed understanding of shelter needs and quality, more accurate assessment of ongoing rescue operations, faster identification of lost or injured people, etc. However, implications of the multimedia content on social media have not yet been studied in depth, unlike their text-only counterparts. There are only a few recent studies in this emerging research area that explore how social media image and video content can provide critical information, especially during crisis events such as natural disasters, for emergency management and response organizations.

*Sifting through social media imagery data for identifying relevant and unique content*

As shown by (Peters and Joao 2015; Chen et al. 2013; Nguyen, Alam, et al. 2017), social media images captured during a crisis event are oftentimes not related to the crisis event itself or do not always present relevant information for emergency management and response organizations. Social media users can post all sorts of images using event-specific hashtags (such as *#HurricaneHarvey*, *#HurricaneIrma*, or *#HurricaneMaria*) to advertise their content, even though this can be considered as unethical behavior during natural disasters and emergencies. Besides relevancy, redundancy in social media images is another important issue that needs to be addressed in order to extract succinct information useful for humanitarian organizations. People can just re-tweet an existing image (i.e., exact duplicates), or share slightly-altered (e.g., rescaled, cropped, text embedded, etc.) versions of an existing image (i.e., near duplicates). These images usually do not provide any additional contextual information, and hence, should be eliminated from the data processing pipeline for optimal use of time as well as human and machine computation resources during crisis situations. In light of the aforementioned studies, we investigate the feasibility of cleaning social media imagery data from irrelevant and redundant content and analyze whether social media imagery can be a source of information for crisis response and management.

*Extracting actionable information for situational awareness*

Detection of relevant and unique multimedia content is certainly necessary but not sufficient in the context of crisis response and management. Humanitarian organizations do need more concise situational awareness information to assess the overall crisis situation. In order to utilize the full potential of multimedia content available on social media, accurate machine learning models should be developed for each particular humanitarian use case. For example, understanding the extent of the infrastructure and utility damage caused by a disaster is one of the core situational awareness tasks listed earlier. Several studies in the literature have already shown that social media images can be analyzed for automatic damage assessment in addition to the textual content analysis (Daly and Thom 2016; Lagerstrom et al. 2016; Liang et al. 2013; Nguyen, Ofli, et al. 2017). Inspired by these studies, we perform an infrastructural damage assessment task on cleaned social media imagery content.

**LITERATURE REVIEW**

Current literature shows the use of social media such as Twitter, Facebook and YouTube for curating, analyzing and summarizing crisis-related information in order to make decisions and responses (Imran, Castillo, Lucas, et al. 2014; Vieweg, Hughes, et al. 2010; Imran, Castillo, Diaz, et al. 2015; Terpstra et al. 2012; Tsou et al. 2017). Among social media studies, most of them focus on Twitter, mainly because of its timeliness and availability of information from a large user base. In (Hagen et al. 2017), the authors analyzed Twitter network structure to understand the flow of information and how different actors and communities contribute towards influential topics. Avvenuti et al. investigate Earthquake Alerts and Report System, which exploits tweets, to understand how such systems can be useful during crisis-related events (Avvenuti et al. 2017). The system collects tweets during an ongoing crisis event, filters irrelevant content, detects an event, assesses damage, and for the sake of comprehensibility, it provides a visualization. Authors conclude that such a system is highly important for crisis-related events. The study of Kim and Hastak investigate how emergency agencies and organizations can better plan operation strategies for a disaster by utilizing individuals' information on a social network (Kim and Hastak 2018).

For the automatic analysis of social media textual and multimedia streams current literature report several AI and computational methods. Most of these methods mainly use supervised or unsupervised (e.g., clustering and topic modeling) techniques. The supervised techniques include classic machine learning algorithms such as Random Forest and Deep Neural Network such as Convolutional Neural Network (CNN), for a complete survey of these techniques and their applications in the crisis informatics domain see (Imran, Castillo, Diaz, et al. 2015; Castillo 2016).

The state-of-the-art research on sentiment analysis is mostly focused on classifying sentiment in either one of two labels (i.e., positive or negative) or five labels (i.e., very positive to very negative) from textual information (Pang, Lee, et al. 2008) such as movie-reviews (Pang, Lee, and Vaithyanathan 2002), tweets (Paltoglou and Thelwall 2010), and newspaper articles and comments (Celli et al. 2016). For sentiment analysis, one of the commonly used approaches is to use sentiment lexicon (i.e., SentiWordNet, Sentiment Treebank, and Psycholinguistic features) (Cambria et al. 2016; Socher et al. 2013; Alam, Danieli, et al. 2018) as features for designing the sentiment classifier. In (Nagy and Stamberger 2012), the authors report the use of emoticons along with SentiWordNet helps in improving the classification of sentiment from microblogs dataset collected during disasters and crises. Socher et al. present the use "Sentiment Treebank" can help in detecting sentiment labels with an accuracy of 80.7% to 85.4% (Socher et al. 2013). Other common approaches include the utilization of word embeddings along with deep neural networks. The extensive comparative studies can be found in SemEval tweet classification task (see Rosenthal et al. 2017). Over time several open-source tools have also been developed. Among them, one of the most widely used tools is the Stanford CoreNLP Natural Language Processing Toolkit (Manning et al. 2014), which supports all preprocessing to sentiment classification methods. In our study, we used Stanford sentiment analysis toolkit, which allowed us to classify tweets in five labels (i.e., very positive to very negative).

For general text classification task, current literature shows the use of classic algorithms such as Maximum Entropy, Logistic Regression, Random Forest, Naïve Bayes classifier, and Conditional Random Fields (CRFs) and deep learning based techniques such as Convolutional Neural Network (CNN) (Nguyen, Al-Mannai, et al. 2017), and Long-Short-Term-Memory (LSTM) (Rosenthal et al. 2017). For tweet classification of this study, we used an in-house developed classifier that uses the Random Forest learning scheme.

Since the supervised approach of text classification requires human annotated labels, the use of the semi-supervised and un-supervised approaches has been increased. For text analysis one of the well-known techniques is topic modeling, which uses Latent Dirichlet Allocation (LDA) (Blei et al. 2003),–a generative probabilistic model. It provides an explicit representation of a textual content. For tweet analysis, there are many studies which analyze tweets to extract information using different variants of LDA topic modeling technique (Mendoza et al. 2010; Chae et al. 2014; Yang et al. 2014; Hong and Davison 2010). Our study of topic modeling is based on LDA.

The extraction of Named Entity Recognition (NER) has a long history in NLP for extracting entities from newspaper articles (Alam, Magnini, et al. 2015) to biomedical text (Alam, Corazza, et al. 2016). The extraction of named entities from Tweets is more challenging due to the noisy structure of the data. Recent approaches for entity recognition from tweets include LSTM, Bidirectional-LSTM, and CRFs (Baldwin et al. 2015; Limsopatham and Collier 2016; He and Sun 2017). For NER task, we used Stanford NER toolkit, which is based on CRFs.

In addition to the textual content analysis, recent studies have also been focused on the analysis of imagery content shared on social media (Nguyen, Alam, et al. 2017; Nguyen, Ofli, et al. 2017; Alam, Imran, et al. 2017). Combining textual and visual content could provide highly relevant information. In (Bica et al. 2017), the authors study the social media images posted during two major earthquakes in Nepal during April and May 2015. Their study focused on identifying geo-tagged images and their associated damage, local vs. global information fusion with images and they also developed an annotation scheme for image analysis. Their findings suggest that global twitter user emphasize recovery and relief effort while local user emphasizes suffering and major damage. In a most recent study, Alam, Ofli, et al. reported an image processing pipeline to extract meaningful information from social media images during a crisis situation, which has been developed using deep learning based techniques (Alam, Ofli, et al. 2018). For this study, we used the same system to execute all of our image processing tasks.

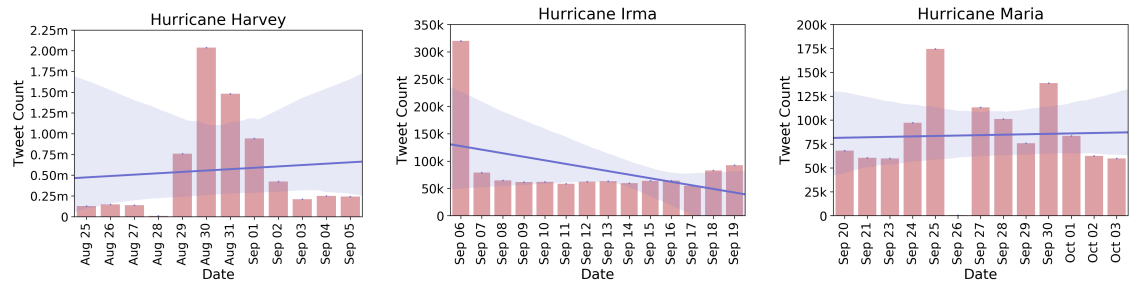

**Figure 1. The total number of tweets collected for each event per day – Harvey (left), Irma (center) and Maria (right). Purple lines indicate trends in the daily tweet data volume.**

In summary, we used several supervised and unsupervised machine learning techniques in this study to analyze three natural disasters. We have conducted a large-scale data analysis comprising of textual and imagery content, which are automatically collected from Twitter. Our analyses include i) finding relevant topic using topic modeling, ii) classifying tweets into humanitarian categories using supervised classification, iii) extracting named entities using NLP techniques, iv) detecting sentiment, v) identifying duplicate images for filtering based on a perceptual hash technique, vi) finding relevant images, and vii) classifying damage severity level from images.

**DATA COLLECTION AND DESCRIPTION**

This study uses Twitter data collected during three natural disasters: Hurricanes Harvey, Irma, and Maria. We used the Twitter streaming API to collect tweets that match with event-specific keywords and hashtags. This API returns 1% of the whole Twitter data at a particular time. Next we describe each dataset in detail.

**Hurricane Harvey**

According to Wikipedia[3], Hurricane Harvey was a category 4 storm when it hit Texas, USA on August 25, 2017. It caused nearly USD 200 billion in damage, which is record-breaking compared with any natural disaster in US history. For Hurricane Harvey, we started the data collection on August 25, 2017 using keywords: *"Hurricane Harvey"*, *"Harvey"*, *"HurricaneHarvey"* and ended on September 5, 2017. In total, 6,732,546 tweets were collected during this period. Figure 1 (left chart) depicts the distribution of daily tweets in this collection. Surprisingly, we have a significantly lower number of tweets (i.e., less than 0.25 million per day) from August 25 to August 28[4] compared to the next five days where a 2-million-tweet peak can be observed on a single day, i.e., August 30. Among the Hurricane Harvey tweet data, 115,525 were found to have an image URL.

**Hurricane Irma**

Hurricane Irma[5] caused catastrophic damage in Barbuda, Saint Barthelemy, Saint Martin, Anguilla, and the Virgin Islands. On Friday September 8, a hurricane warning was issued for the Florida Keys and the Florida governor ordered all public schools and colleges to be closed. The Irma storm was a category 5 hurricane, which caused USD 66.77 billion in damage. We collected Hurricane Irma-related data from Twitter starting from September 6, 2017 to September 19, 2017 using the keywords: *"Hurricane Irma"*, *"Irma storm"*, *"Irma"*. In total, 1,207,272 tweets were collected during this period. Figure 1 (middle chart) shows the distribution of daily tweets of Hurricane Irma data. On the first day (i.e., September 6), we can see a surge of tweets in which more than 300,000 tweets were collected. However, during the next days the distribution stayed steady around 50,000 tweets per day. Besides, 60,973 of these Hurricane Irma tweet data contained image URLs.

---

[3]https://en.wikipedia.org/wiki/Hurricane_Harvey
[4]On August 28, due to a network issue, only 9,825 tweets were collected.
[5]https://en.wikipedia.org/wiki/Hurricane_Irma

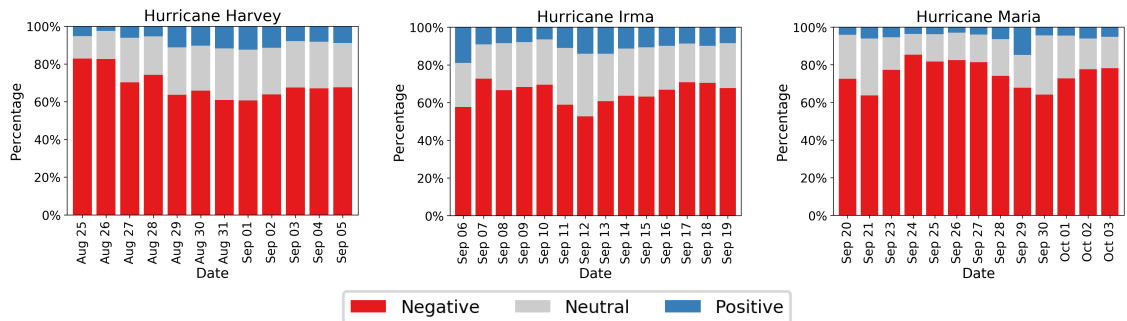

**Figure 2.** Sentiment analysis results: distribution of daily sentiment for Harvey (left), Irma (center), and Maria (right).

**Hurricane Maria**

Hurricane Maria[6], was a category 5 hurricane that slammed Dominica and Puerto Rico and caused more than 78 deaths including 30 in Dominica and 34 in Puerto Rico, while many more left without homes, electricity, food, and drinking water. The data collection for Hurricane Maria was started on September 20, 2017 and ended on October 3, 2017. In total, we collected 1,096,335 tweets during this period using keywords: *"Hurricane Maria", "HurricaneMaria", "Tropical Storm Maria", "Maria Storm"*. The right chart in Figure 1 shows the daily tweet distribution for the Hurricane Maria data. Of these Hurricane Maria tweet data, we found 19,681 tweets with image URLs.

We note that the keywords selected to collect data are high-level but specific to the hurricanes events. The Twitter API returns tweets that mention these keywords. However, high-level keywords could bring more noisy data compared to specific keywords. This is due to the irrelevant messages that people post about mundane events, advertisements, jokes, etc.

## EXPERIMENTS, RESULTS, AND DISCUSSION

To analyze the three disaster events from the social media lens and to fulfill the objectives set in the objectives section, we perform an extensive experimentation using the collected datasets. This section provides details of our experimental setup and the analysis results.

### Preprocessing

Since the tweet texts are noisy, preprocessing is required before using them in further analysis. The processing part of this study includes removal of stop words, non-ASCII characters, punctuations (replaced with whitespace), numbers, URLs, and hastags.

### Sentiment analysis

To perform the sentiment analysis, we used the Stanford sentiment analysis classifier (Socher et al. 2013) to extract the sentiment labels that are being expressed or semantically represented in the tweets. The Stanford sentiment analysis classifier consists of 5 categorical labels such as *Very Negative*, *Negative*, *Neutral*, *Positive* and *Very Positive*. For each tweet the classifier assigns one of the five categories with their confidence. The classifier has been designed using Recursive Neural Tensor Network and exploiting sentiment treebank, which consists of fine-grained sentiment labels for 215, 154 *phrases* in the parse trees of 11, 855 *sentences*. The accuracy of the classifier for fine-grained sentiment labels is 80.7% as presented in (Socher et al. 2013). For our task, we fed the preprocessed tweet to the classifier to get the classifier's prediction. For the sake of presenting the results, we combined "Positive" and "Very positive" classes into a single class "Positive", and the "Negative" and "Very negative" classes into a single class "Negative".

In Figure 2, we present the distribution of sentiment results in terms of three classes for each day for the three events. One can clearly observe that the "Negative" sentiment dominates for all the three events throughout all days. We also observe that one of the factors that cause high negative sentiment in the early stage of a disaster is due to the use of aggressive language, e.g., using "F**k", "s**t", "b***h" like words, cursing disaster, complaining, and expressing anger when someone's plan gets disturbed due to the disaster event. However, negative sentiment during the event is mainly caused by actual issues faced by affected people where there is no response from the government or concerned authorities or the response is slow. For example, this includes cases like a power outage, lack of drinking water, food, or other necessary facilities. Real-time monitoring of sentiment prediction of social media streams can be useful for the concerned response authorities to understand people's sentiment and concerns, and to guide their response efforts towards addressing those concerns.

---

[6] https://en.wikipedia.org/wiki/Hurricane_Maria

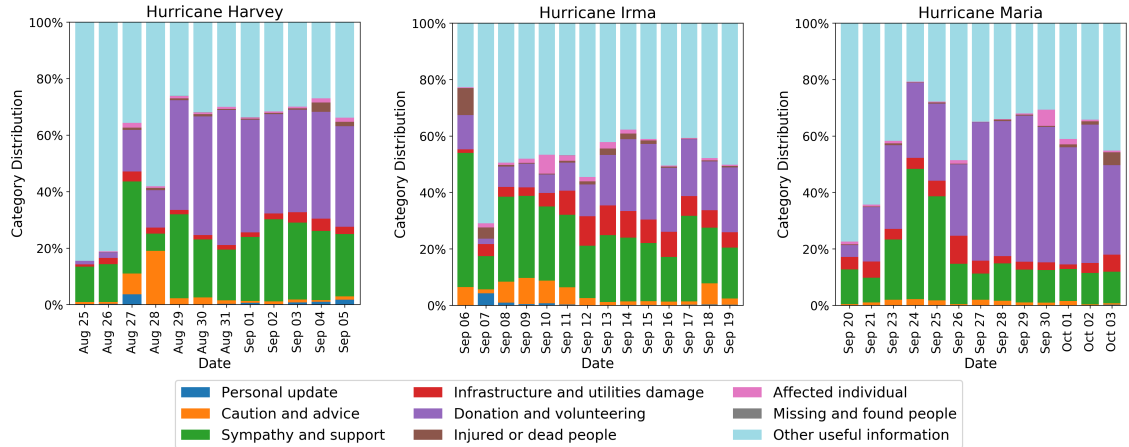

**Figure 3.** Temporal distribution of categories based on the automatic classification of tweet text into one of the humanitarian categories for Harvey (left), Irma (center) and Maria (right).

**Table 1. Data for Humanitarian Categories Classification**

| **Classes** | **Train (60%)** | **Dev (20%)** | **Test (20%)** |
|---|---|---|---|
| *Affected individual* | 3029 | 757 | 758 |
| *Caution and advice* | 3288 | 822 | 822 |
| *Donation and volunteering* | 4278 | 1070 | 1070 |
| *Infrastructure and utilities damage* | 3189 | 797 | 798 |
| *Injured or dead people* | 2148 | 537 | 538 |
| *Missing and found people* | 405 | 101 | 102 |
| *Personal updates* | 968 | 242 | 242 |
| *Other useful information* | 4000 | 2000 | 2000 |
| *Sympathy and support* | 5504 | 1376 | 1376 |
| *Not related or irrelevant* | 4000 | 2000 | 2000 |
| **Total** | 30809 | 9702 | 9706 |

**Classification of humanitarian categories**

When information needs are known, as in the case of many humanitarian organizations, using supervised machine learning techniques is preferred over unsupervised techniques. In this section, we report the results obtained from the supervised classification of the three events' data. For this purpose, we used a decision tree based learning scheme known as Random Forest. To train the supervised machine learning model, we used human-labeled Twitter data from a number of past disasters (Imran, Mitra, and Castillo 2016). The labeled data were collected during more than 30 past disasters including hurricanes, earthquakes, and floods.

Table 1 shows the class distribution of our train, development (dev), and test sets. The train set is used to learn the model, the development (dev) set is used for parameter-tuning, and the test set is used for the model evaluation. To learn the model, we used a bag-of-words approach. The performance obtained using the test set in terms of the F-measure is $F1 = 0.64$ and accuracy of 0.66. In Table 2, we also present the results of individual classes. The trained model is then used to classify all the tweets of the three events.

In Figure 3, we present day-wise distribution of the automatically classified tweets for the three events. One clear pattern representing one of the most prevalent categories "Other useful information" across all the events. The second most dominant category is "Donation and volunteering". This category contains tweets either requesting any kind of donations, or offering help or donations. However, the donation category seems to emerge slowly as the disaster event progresses. For instance, see the first few days of all three events when fewer donation-related data is found. The "Sympathy and support" category contains prayers and thoughts messages and it seems consistent across all the events and days except the last five days of the hurricane Maria event in which we observe the lower proportion of sympathy messages compared to the other two events. Caution and advice messages seem to appear mostly at the beginning of an event and slowly dispersed. The "Infrastructure and utilities damage" category gradually emerge and gets more prominent during the middle days of the events. The "Injured or dead people" category appears during some days of Irma and Maria, but the "Missing and found people" was among minority, as actually there were no missing or found cases reported during the three events. To compare the proportion of the informative categories with "Irrelevant or not related" category, in Figure 4 we show the distribution of Relevant versus Irrelevant messages for each day of the three events. The relevant category, in this case, is a combined version of all the informative categories described above. The proportion of relevant messages seems to follow a decreasing trend from 80% to 60% during Harvey unfolding while it looks approximately constant around 60% for Irma and 70% for Maria.

**Table 2. Text classifier performance in terms of Precision (P), Recall (R), and F1-score.**

| Class | P | R | F1 |
|---|---|---|---|
| **Affected individual** | 0.70 | 0.72 | 0.71 |
| **Caution and advice** | 0.63 | 0.64 | 0.64 |
| **Donation and volunteering** | 0.69 | 0.80 | 0.74 |
| **Infrastructure and utilities damage** | 0.65 | 0.64 | 0.65 |
| **Injured or dead people** | 0.85 | 0.87 | 0.86 |
| **Missing and found people** | 0.64 | 0.21 | 0.31 |
| **Not related or irrelevant** | 0.68 | 0.73 | 0.71 |
| **Personal updates** | 0.66 | 0.64 | 0.65 |
| **Relevant information** | 0.68 | 0.40 | 0.50 |
| **Sympathy and support** | 0.55 | 0.77 | 0.64 |

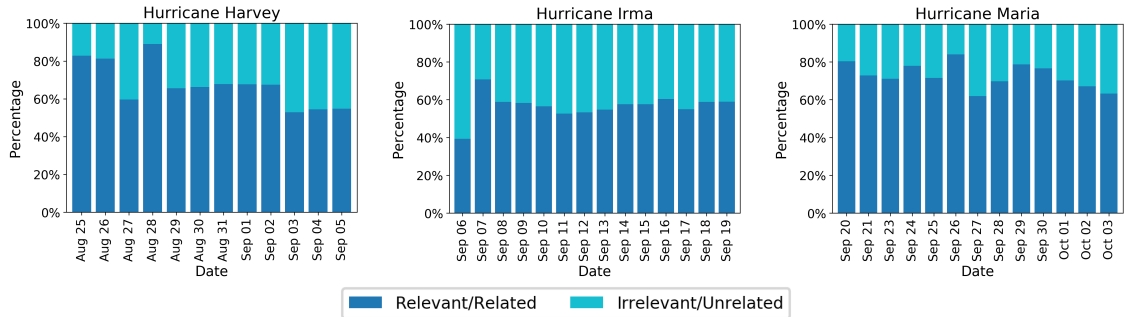

**Figure 4. Daily proportion of unrelated or irrelevant tweets to related or relevant ones for Harvey (left), Irma (center) and Maria (right).**

**Topic modeling**

To understand the topics of discussion on Twitter during the three disasters, we used LDA (Blei et al. 2003), which is a well-known topic modeling technique, to generate topics from large amounts of textual data. Topic modeling helps understand and summarize large collections of textual information. With topic modeling, we aim to discover hidden topical patterns that are present across different days of a disaster event. Moreover, with human intervention, we aim to analyze these topics and annotate them to summarize a discussion point. We apply LDA to generate 10 topics from the preprocessed tweets using each day's data from all three events. Due to space limitations, we do not present all the results of 10 topics for each day, instead, we show the top 30 most relevant words (i.e., words with a high probability of being associated to a topic) from the most prevalent topic among the 10 topics obtained from randomly selected four days of an event.

Figure 5 depicts the results obtained from the Hurricane Harvey data. Figure 6 shows the results obtained from the Hurricane Irma data and Figure 7 shows the topics results obtained from the Hurricane Maria data. Different than the traditional clustering techniques in which one data point (e.g., the text of a tweet in our case) can only belong to one cluster/topic, in topic modeling, it can belong to different topics. For example, a tweet can talk about different aid needs like food, water, shelter. For this reason, in the results, you might notice same words appear in multiple topics. The red bars indicate the word/term frequency for a given topic. The blue bars show the term frequency for a given day (i.e., how many times a word appear in a given day). All the words shown in a figure belong to one topic for a given day. For instance, the most prevalent topic emerged on August 31 during the Hurricane Harvey contains "evacuees", "mattress", and "redneck" as the top three most relevant words. The reason behind this topic was about a mattress chain owner who offered his stores for Harvey evacuees and his trucks for rescue operations. Similarly, on September 4 during the Hurricane Harvey, one of the discussion points was about "rescuing", "paramedic", "worked", as shown in Figure 5. Upon investigation, it appeared that the topic emerged due to a large number of tweets about a paramedic who reportedly worked six straight days to rescue Harvey victims, so the discussion built on the topic that he may be deported if Trump ends DACA.

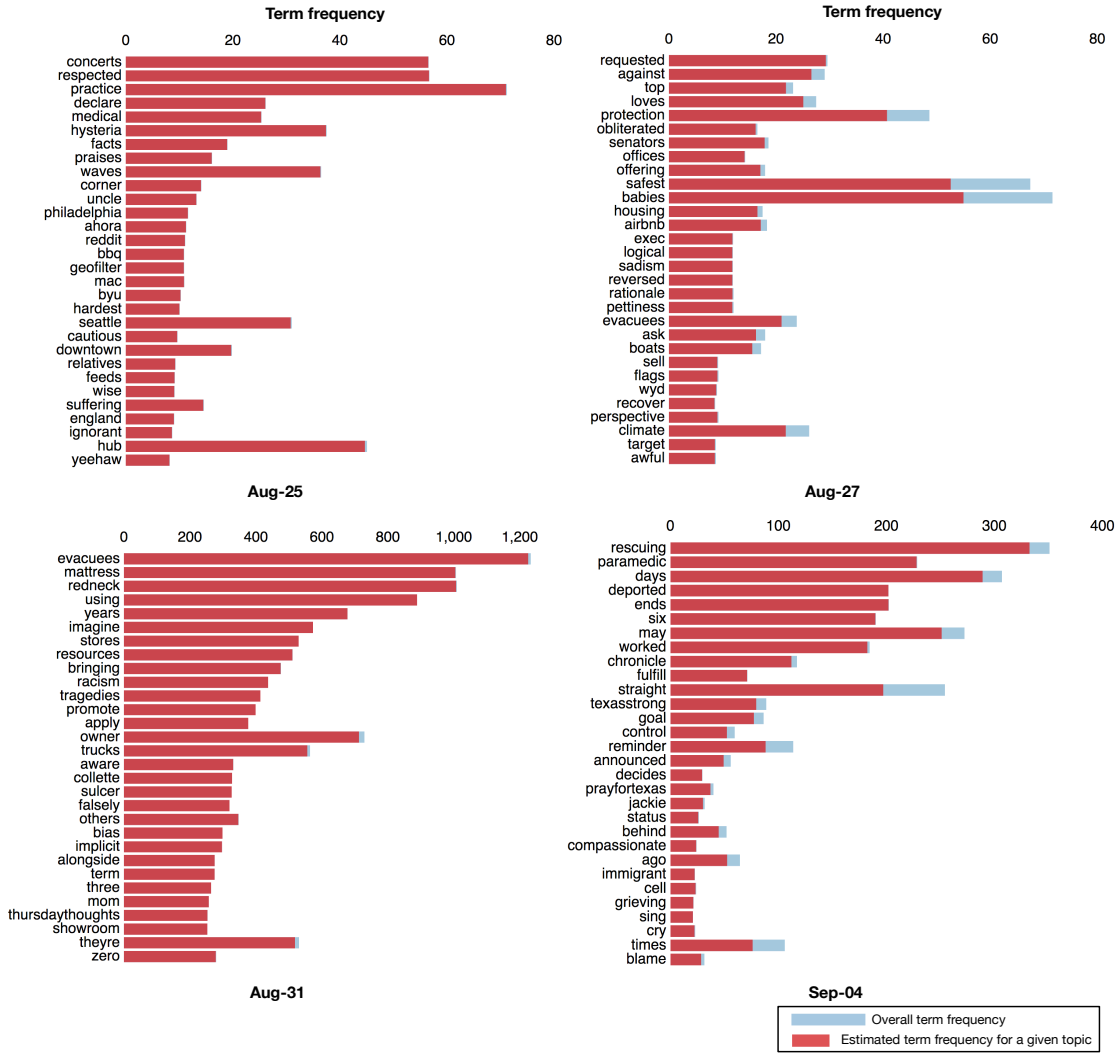

**Figure 5.** LDA generated topics from some selected days of Hurricane Harvey. We show top 30 most relevant words from the most prevalent topic among the 10 topics from a given day.

One of the dominant topics that emerged on September 6 during the Hurricane Irma contains words: "help", "years", "supplies". This was due to a large discussion about *"Trump plans to ship 800,000 Dreamers back to Mexico...Mexico ships 25 trailers of supplies to help Americans..."*. Among the topics obtained from the Hurricane Maria data, there is a topic about people's concern regarding *"There will be no food in Puerto Rico. There is no more agriculture in Puerto Rico for years"*. However, on October 2, an important topic emerged about some strike of truck drivers in Puerto Rico delaying donation goods delivery. Overall, we observed the LDA generated topics could reveal critical public issues and concerns during disasters. For response organizations, anticipating such issues is hard. However, a system that identifies prominent and emerging topics over time (e.g., per day or per 6-hour period) to inform emergency managers about important issues that public is facing would be hugely helpful.

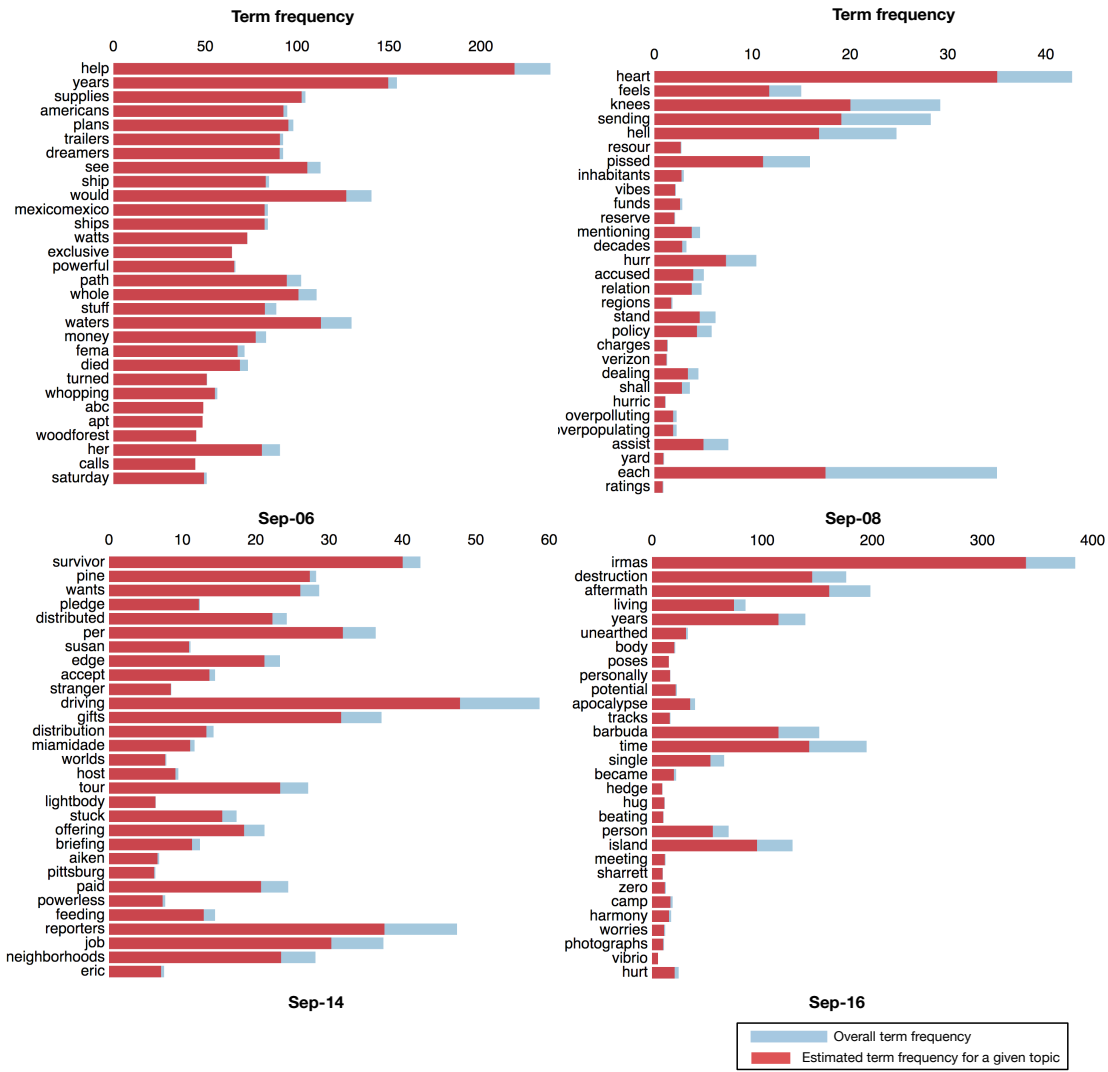

**Figure 6. LDA generated topics from some selected days of Hurricane Irma. We show top 30 most relevant words from the most prevalent topic among the 10 topics from a given day.**

**Named entity recognition**

Named entity recognition has been proposed as part of a visual analytic tool in (Aupetit and Imran 2017) to extract top-k entities from tweets. We follow the same approach here, using the Stanford NER toolkit (Finkel et al. 2005) to analyze the three hurricanes in terms of top "Persons", "Organizations" and "Locations" mentioned in collected tweets. The reported F-measure of this NER system is 86.72% to 92.28% for different datasets.

Regarding Hurricane Harvey (Table 3), "Steve Harvey" is the most mentioned person with 7,014 tweets. Steve Harvey is an American comedian hosting "The Steve Harvey Morning Show" on a Los Angeles (CA) based radio. On August 24, Twitter user @*zwearts* tweeted a photoshopped picture of the hurricane with comedian Steve Harvey's face in the center[7], which soon got viral. We report that almost all the tweets associated with this particular entity are not helpful for disaster response, so a responder can simply ignore these ~7k tweets. The entity "EPA" of type "Organizations" is the 5th most mentioned organization with 2,828 tweets from 1,959 unique messages.

---

[7] http://knowyourmeme.com/memes/events/2017-hurricane-harvey

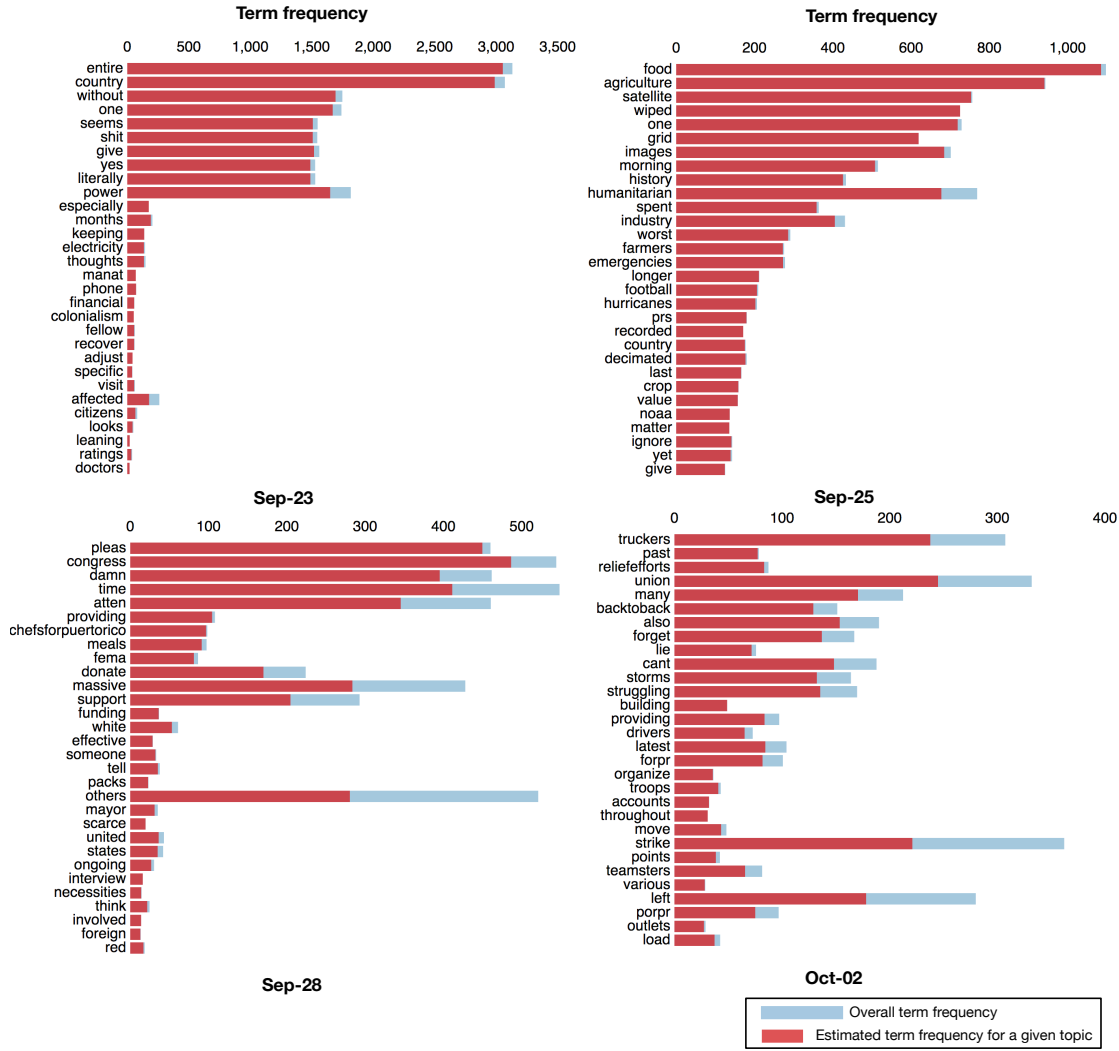

**Figure 7. LDA generated topics from some selected days of Hurricane Maria. We show top 30 most relevant words from the most prevalent topic among the 10 topics from a given day.**

EPA is the United States Environmental Protection Agency. Tweets relate lab testing of contaminated water and increased funding of the EPA due to Harvey event. The "Location" type entity "Oregon" is mentioned 1,087 times. Messages complain about Harvey and then announced Irma attracting more attention and funds while wildfire ravaged forests in Oregon[8]. Other tweets mentioning "Oregon" talk about the unexpectedly high number of natural disasters (Harvey, Irma and wildfires) occurring at the same time.

During Hurricane Irma (Table 4), the entity "Limbaugh" was the second most mentioned person's name with 1,918 tweets. Rush Limbaugh is a host of a Florida-based radio talk show, who angered many people by discussing that media and meteorologists were exaggerating the threat from Hurricane Irma to comply with a political agenda regarding the effect of the climate change. He eventually evacuated Palm Beach himself on September 8[9]. Another entity "Tim Duncan", a former NBA player, is ranked 2nd with 1,460 mentioning tweets. He launched a fund raising campaign starting on the 9th September for the U.S Virgin Islands where he was born and grew up[10]. "Trump Hotels" appear as the 2nd top cited organization with 1,500 tweets, following news reports[11] that some hotels, which are the properties of U.S. President Donald Trump might be hit by Hurricane Irma. It turned out that the same mes-sage was retweeted multiple times: **"RT @billprady: Correcting a rumor: Trump Hotels in FL are NOT housing displaced persons from Irma. Do not call 1.855.878.6700 all day long"**. Still filtering out redundant messages, 25 unique messages related to "Trump hotels" were talking about the same fake news. Top-10 locations that are found from Irma data have no particular interest.

---

[8]https://www.usatoday.com/story/opinion/2017/09/20/wildfire-crisis-has-been-blanketed-hurricane-hysteria-ken-fisher-bruce-658959001/
[9]https://www.washingtonpost.com/news/the-fix/wp/2017/09/06/rush-limbaughs-dangerous-suggestion-that-hurricane-irma-is-fak
[10]https://www.theplayerstribune.com/tim-duncan-hurricane-irma-us-virgin-islands
[11]http://time.com/4931642/donald-trump-hurricane-irma-properties/

| Persons | | Organizations | | Locations | |
|---|---|---|---|---|---|
| **Steve Harvey** | **7,014** | Congress | 15,302 | Houston | 58,731 |
| Donald Trump | 5,181 | FEMA Relief Funds | 5,304 | Texas | 47,582 |
| Obama | 4,888 | Coast Guard | 4,111 | Mexico | 18,107 |
| Jackie | 4,719 | American Red Cross | 3,215 | Florida | 10,598 |
| Bush | 2,502 | **EPA** | **2,828** | United States | 3,579 |
| Hillary | 2,053 | DACA | 2,490 | America | 2,245 |
| Mike | 2,018 | GOP | 2,000 | Haiti | 1,435 |
| Clinton | 1,974 | Justice League | 1,897 | U.S. | 1,408 |
| Alex Jones | 1,763 | Red Cross | 1,819 | Fort Lauderdale | 1,176 |
| Alonso Guillen | 1,644 | Republican Party | 1,708 | **Oregon** | **1,087** |

**Table 3. Hurricane Harvey top-10 entities. Numbers represent the amount of tweets mentioning the entities. In bold font, the cases discussed in the paper.**

| Persons | | Organizations | | Locations | |
|---|---|---|---|---|---|
| Bryan | 4,427 | Congress | 2,583 | Florida | 61,195 |
| **Rush Limbaugh** | **1,918** | **Trump Hotels** | **1,500** | Miami | 14,876 |
| **Tim Duncan** | **1,460** | Coast Guard | 886 | Caribbean | 12,169 |
| Rick Scott | 1,362 | CNN | 866 | Cuba | 6,159 |
| Marc Bell | 848 | American Red Cross | 594 | Virgin Islands | 5,878 |
| Donald Trump | 784 | NASA | 537 | Bahamas | 4,880 |
| Hillary | 542 | NYT | 535 | US | 3,604 |
| Katie | 465 | National Hurricane Center | 521 | Long Island | 3,083 |
| Robert De Niro | 430 | Red Cross | 495 | South Florida | 2,868 |
| Tim Tebow | 386 | BBC News | 477 | Orlando | 2,691 |

**Table 4. Hurricane Irma top-10 entities. Numbers represent the amount of tweets mentioning the entities. In bold font, the cases discussed in the paper.**

During Hurricane Maria (Table 5) the entity mention "2nd Class Brandon Larnard", a Naval aircrewman, was part of the military hurricane response efforts the Pentagon deployed on the island of Dominica and Puerto Rico. His picture[12] showing him carrying an evacuee toddler was retweeted 579 times putting his name in 2nd position of mentioned person's names. The organization entity "AOKBOKCreative" was top 3 organization cumulating 1, 075 retweets after a message[13] claiming they raised 500K USD in 90 minutes to support Hurricane Maria victims. The location entity "Vietnam" is mentioned in 309 messages where it relates to a Vietnam veteran claiming Hurricane Maria was worse than war[14].

Overall, our study suggests that the identification of most mentioned named entities can help discover important stories either to filter them out ("Steve Harvey" case during hurricane Harvey) in order to focus on messages related to actual local emergency needs, or to consider some messages for further detailed analysis ("Oregon" case during Hurricane Harvey). Our analysis also illustrates the complexity of the named entity recognition problem, for instance "Harvey", "Irma", "Jose", "Katie", are all hurricane names detected as persons' names that require manual checking of the texts in tweets to decide about them. In other cases, "US" can be considered an organization or a location depending on the context. Therefore manual editing was necessary to correct for the automatically discovered named entities, emphasizing the needs for an interactive integrated tool, such as the one proposed in (Aupetit and Imran 2017).

---

[12]https://www.voanews.com/a/pentagon-names-three-star-general-to-head-up-puerto-rico-relief/4048825.html
[13]https://twitter.com/search?q=%23AOKBOKCreative&src=typd&lang=en
[14]http://edition.cnn.com/2017/09/25/us/puerto-rico-hurricane-maria-combat-veteran/index.html

| Persons | | Organizations | | Locations | |
|---|---|---|---|---|---|
| Donald Trump | 2,909 | Congress | 7,295 | Puerto Rico | 42,312 |
| **2nd Class Brandon Larnard** | **579** | NFL | 1,218 | Virgin Islands | 8,806 |
| Carmen Yulin Cruz | 443 | **AOKBOKCreative** | **1,075** | San Juan | 3,354 |
| John McCain | 343 | National Guard | 603 | US | 3,187 |
| Luis Fonsi | 306 | White House | 492 | United States | 2,004 |
| Elaine Duke | 271 | GOP | 402 | Caribbean | 816 |
| Taylor | 199 | Army Corps | 304 | Puerto Ricans | 551 |
| Patty | 189 | DHS | 288 | Florida | 464 |
| Bush | 152 | Teamsters | 268 | Russia | 452 |
| Levine | 120 | NYT | 234 | **Vietnam** | **309** |

**Table 5. Hurricane Maria top-10 entities. Numbers represent amount of tweets mentioning the entities. In bold font, the cases discussed in the paper.**

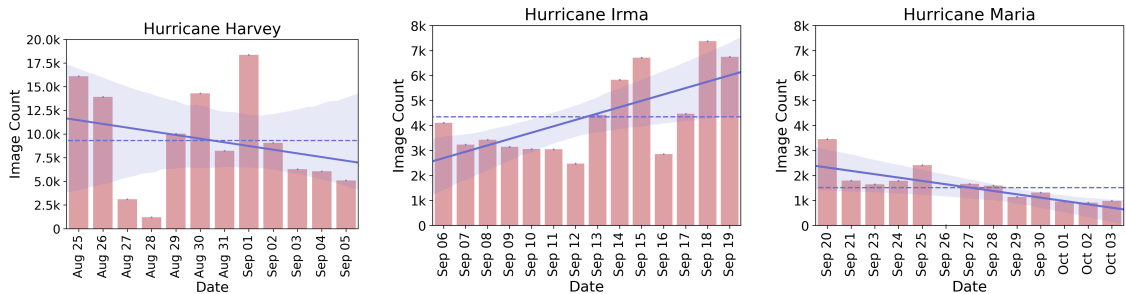

**Figure 8. Bar plot of total number of image tweets per day for Harvey (left), Irma (center), and Maria (right). Horizontal dashed lines show the *average* number of image tweets per day whereas the solid lines indicate the changing trends in the volume of social media imagery data.**

**Multimedia content analysis**

As discussed earlier, images convey more information about the disaster situation than simple words, and hence, analysis of social media image data bears significant potential for crisis response and management purposes. A reasonable proportion of tweets posted during natural disasters include imagery content. Specifically, Hurricane Harvey data include a total of 110,597 images, whereas Hurricane Irma and Maria data include a total of 60,932 and 19,681 images, respectively. Figure 8 shows the distribution of image tweets collected on each day for each event. On average, daily volume of image tweets is higher for Hurricane Harvey (i.e., ∼9.3k) than those of Hurricane Irma (i.e., ∼4.3k) and Hurricane Maria (i.e., ∼1.5k). While the total number of image tweets per day exceeds 15k on certain days during Hurricane Harvey, the highest number of image tweets remains ∼7k per day for Hurricane Irma and ∼3k per day for Hurricane Maria. For Hurricanes Harvey and Maria, the trend lines indicate a decrease in the total number of image tweets per day as time passes, whereas we see an increase in total number of image tweets per day for Hurricane Irma.

For a detailed analysis of the imagery content, we employed our image processing pipeline presented in (Alam, Ofli, et al. 2018), which has readily available image classification models for relevancy filtering and de-duplication of images as described in (Nguyen, Alam, et al. 2017) as well as a model for damage severity assessment from images as studied in (Nguyen, Ofli, et al. 2017). For the sake of completeness, we next provide brief descriptions of these models and recap their performance scores. Relevancy filtering model operates like a "junk filter" that isolates images showing cartoons, banners, advertisements, celebrities, etc., which are deemed as *irrelevant* content by many humanitarian organizations during disasters. We trained this model as a binary (i.e., relevant/irrelevant) classification task using a transfer learning approach where we fine-tuned a pre-trained deep image recognition model for the relevancy filtering task at hand. On a held-out test set, the resulting model achieved 99% precision and 97% recall. De-duplication filtering model aims to identify exact- or near-duplicate images with little modifications such as cropping/resizing, padding background, changing intensity, embedding text, etc. We implemented this model using a perceptual hash technique to determine whether a given image is an exact- or near-duplicate of previously seen images. The similarity threshold was determined as the optimal operation point of the ROC curve, which yielded ∼90% precision and recall. The damage assessment model categorizes the severity of damage observed in an image into three categories, i.e., *severe*, *mild* or *none*. We trained this three-class classification model using ground truth disaster images annotated by humans following the same transfer learning strategy that we used for our relevancy filtering model. The overall accuracy of the resulting damage assessment models varied from 76% to 90% on held-out test sets depending on the disaster type.

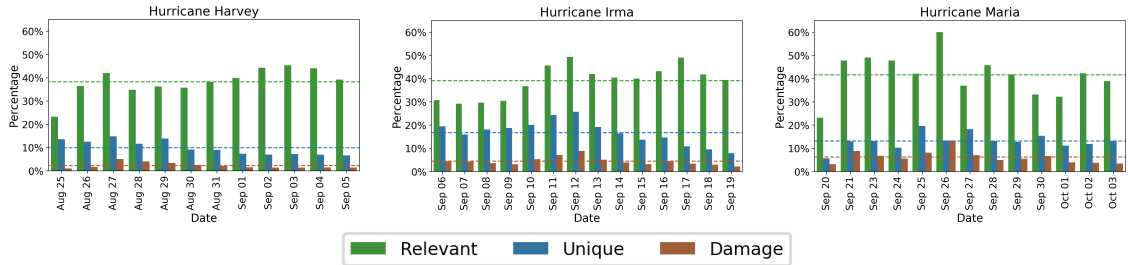

**Figure 9.** Ratio of daily images retained after relevancy filtering (green), de-duplication (blue), and damage assessment (brown) for Harvey (left), Irma (center) and Maria (right).

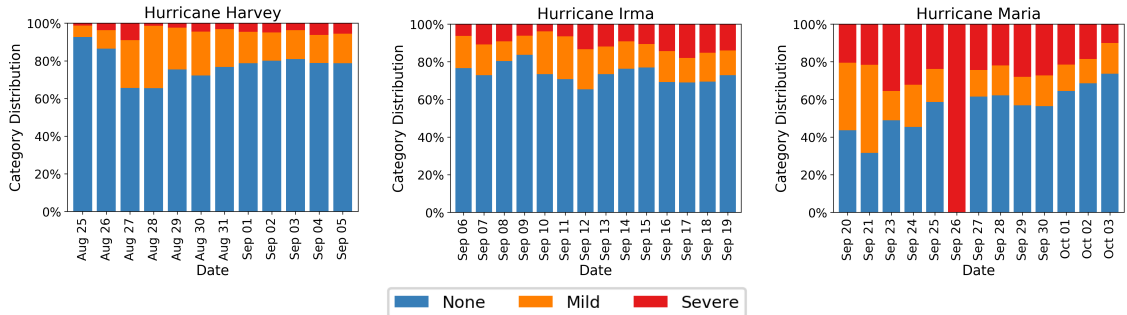

**Figure 10.** Distribution of *severe*, *mild*, and *none* damage images after relevancy and uniqueness filtering for Harvey (left), Irma (center), and Maria (right).

We first analyze the proportions of relevant and unique images in social media imagery data collected during the Hurricanes Harvey, Irma and Maria. The results of our relevancy and uniqueness analyses are in conformity with the past observations of (Nguyen, Alam, et al. 2017) as illustrated by Figure 9. Specifically, around 40% of the images are deemed relevant to the actual disaster event in all Hurricanes Harvey, Irma and Maria, i.e., ~38%, ~39% and ~41%, respectively. Furthermore, Hurricane Irma data contain ~17% unique images whereas Hurricane Maria data contain ~11% and Hurricane Harvey data contain ~10% unique images on average. When looked at carefully, the ratios of both relevant and unique images are relatively higher in the later days of all hurricanes.

Next, we analyze the severity of damage (i.e., *none*, *mild* or *severe*) observed in the set of images that were deemed relevant and unique. Brown bars in Figure 9 indicate the percentage of images with some damage content (i.e., *mild* or *severe*). On average, ~2.5% of Hurricane Harvey images show damage content whereas ~4.4% of Hurricane Irma and ~6.2% of Hurricane Maria images show damage content. Moreover, this ratio can be twice as high in the later days of all disasters. Overall, an interesting observation is that even though the total volume as well as the daily volume of image tweets is relatively smaller for Hurricane Maria, the proportions of unique or damage images are higher than those for Hurricanes Harvey and Irma. Even though daily changes in prevalence of relevant and damage images during Hurricane Maria seem to be strongly correlated ($r = 0.71$, $p < 0.01$), we do not observe a statistically significant correlation between relevant and damage image tweet distributions for Hurricane Irma ($r = 0.41$, $p = 0.14$) and Hurricane Harvey ($r = 0.04$, $p = 0.90$). On the other hand, daily changes in prevalence of unique and damage images during Hurricane Irma seem to be very strongly correlated ($r = 0.85$, $p < 0.001$) whereas they seem to be only strongly correlated ($r = 0.62$, $p < 0.05$) during Hurricane Harvey. Even though we observe a moderate correlation between unique and damage image tweet distributions for Hurricane Maria, this correlation is not statistically significant ($r = 0.44$, $p = 0.13$).

In Figure 10, we take a closer look at the damage assessment analysis of images after relevancy and uniqueness filtering for all events. On any given day, only 20-30% of Hurricane Irma images that are relevant and unique depict some damage content (i.e., *mild* or *severe*), whereas this ratio varies between 30-60% for Hurricane Maria. Furthermore, among those Hurricane Irma images that depict some damage content, we see more examples of *mild* damage than *severe* damage. On the contrary, images with damage content in Hurricane Maria data show more *severe* damage than *mild* damage. Among the three hurricanes, Hurricane Harvey data seems to contain the least proportion of damage content.

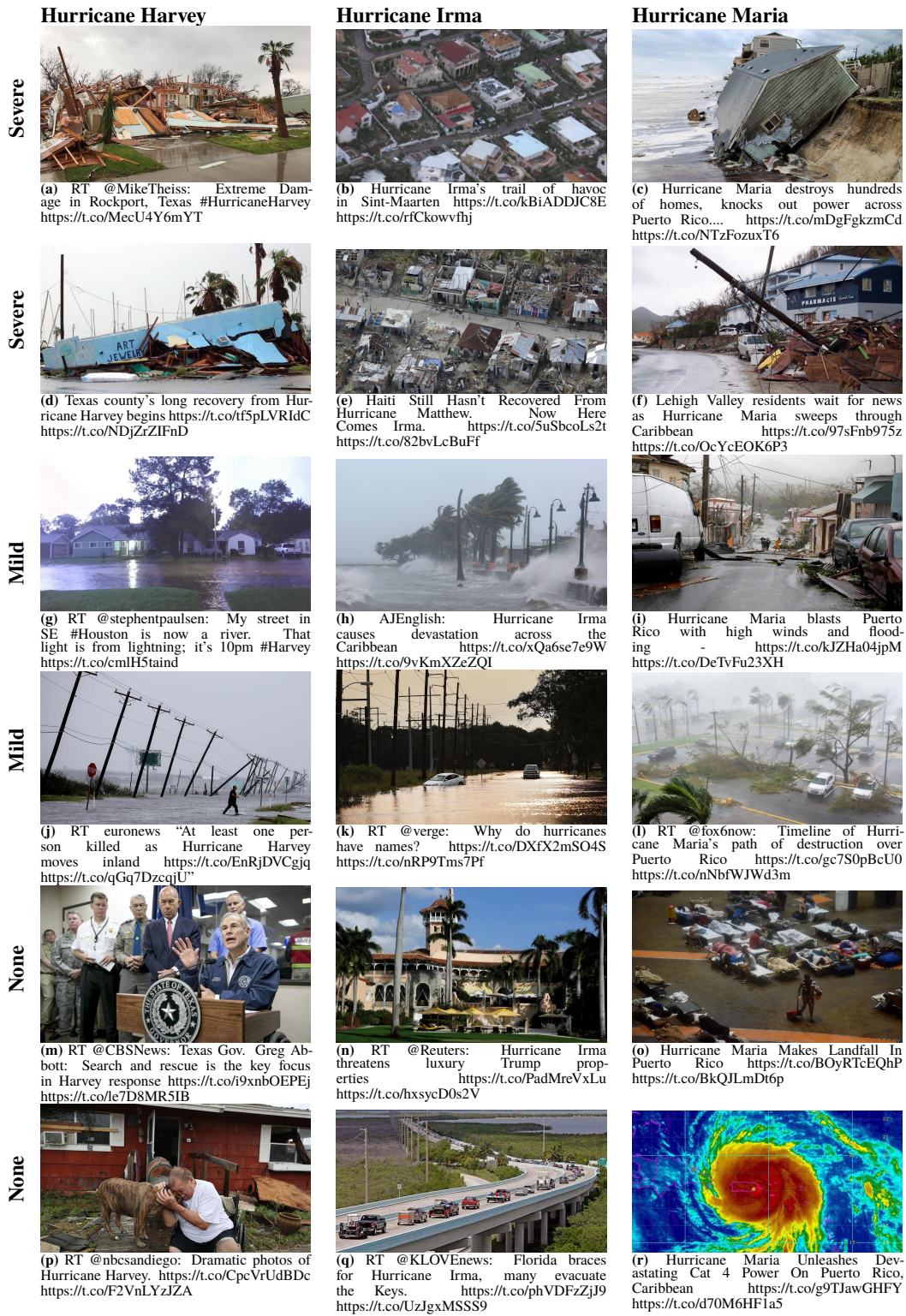

**Figure 11. Sample images with different damage levels from different disaster events.**

Figure 11 shows several example tweets with images classified into different damage categories. In some cases such as Figures 11(a)-(c) and (g)-(i), both tweet text and image content provide critical information related to infrastructure and utilities damage at different severity levels. However, unlike their image counterparts, tweet text in Figures 11(d)-(f) do not provide any significant information about the severity or extent of damage incurred by disaster events other than just stating that the disasters caused some damage. Similarly for tweets in Figures 11(j)-(l), images provide some crucial damage information related to power lines, roads, etc. whereas the corresponding text reports a dead person, questions why hurricanes are named, or mentions the path of the hurricane. More importantly, even though the tweets in Figures 11(m)-(r) do not show any damage content, they provide critical information for other humanitarian categories taxonomy. For instance, Figure 11(o) provides valuable insight for the quality of shelter. Similarly, Figure 11(q) illustrates an example of evacuation and displaced people.

**CONCLUSION**

The widespread use of social media during disasters and emergencies has created opportunities for disaster responders to find useful information for crisis response and management. In this work, we performed an extensive analysis of the data collected from Twitter during three natural disasters. We demonstrated the richness of the useful information both textual as well as multimedia contained in the tweets. By using a wide range of state-of-the-art machine learning techniques, we analyzed millions of tweets and tens of thousands of images to show the utility of the data and potential of the employed techniques for crisis management and emergency response. However, to make sense of the large amounts of crisis-related data on social media, especially during an on-going disaster is still challenging and requires efficient computational methods which can operate in real-time. Among other issues, when it comes to operational crisis management, addressing scalability issues of most of the techniques to process real-time data streams is essential. Furthermore, systems that include humans-in-the-loop for machine training need to deal with humans' limited processing capabilities to maintain high throughput. We consider this as a future work for us and for the crisis informatics community.

**ACKNOWLEDGMENTS**

We would like to extend our sincere thanks to Hemant Purohit from George Mason University and Kiran Zahra from University of Zurich for helping us with the data collection task. We also would like to thank our anonymous reviewers for their detailed and constructive comments.